\documentclass[a4paper]{jpconf}
\newcommand{\comments}[1]{}
\newcommand{\VBFNLO}{{\tt VBFNLO}}
\usepackage{graphicx,amsmath,amssymb}
\begin{document}
\title{Higgs boson production in association with a photon via weak boson fusion 
\footnote{Presented by K.~Arnold at {\em Photon 2011}, Belgium, May 2011}
}
\bibliographystyle{iopart-num}

\clubpenalty=10000
\widowpenalty=10000
\raggedbottom

\author{K.~Arnold$^a$, T.~Figy$^b$, B.~J\"ager$^c$ and D.~Zeppenfeld$^a$\\}

\address{$^a$ Institute for Theoretical Physics, Karlsruhe Institute of Technology, \\ 76128~Karlsruhe, Germany}
\address{$^b$ CERN, CH--1211 Geneva 23, Switzerland \\ \textit{Current affiliation:} School of Physics \& Astronomy, University of Manchester, \\ Manchester M13 9PL, U.K.}
\address{$^c$ Institut f\"ur Physik (THEP), Johannes-Gutenberg-Universit\"at, 55099 Mainz, Germany}

\ead{arnold@particle.uni-karlsruhe.de, terrance.maynard.figy@cern.ch, jaegerba@uni-mainz.de, dieter@particle.uni-karlsruhe.de}

\begin{abstract}
We present next-to-leading order QCD corrections to Higgs production in association with a photon via weak boson fusion at a hadron collider. Utilizing the fully flexible parton level Monte-Carlo program \VBFNLO, we find small overall corrections, while the shape of some distributions is sensitive to radiative contributions in certain regions of phase-space. Residual scale uncertainties at next-to-leading order are at the few-percent level. Being perturbatively well under control and exhibiting kinematic features that allow to distinguish it from potential backgrounds, this process can serve as a valuable source of information on the $Hb\bar{b}$ Yukawa coupling. 

\end{abstract}

\section{Introduction}
Even more than forty years after the incorporation of the Higgs mechanism into the theory of electroweak interactions by Glashow and Salam, its key ingredient, the Higgs boson, still awaits discovery. In this context, physicists around the world hope for the Large Hadron Collider (LHC) to shed light on the nature of symmetry breaking. 

A particularly promising class of reactions in the context of the search for the Higgs boson is provided by weak vector boson fusion processes (WBF). Higgs production via WBF mainly proceeds via the scattering of two quarks by the exchange of massive gauge bosons in the $t$-channel which in turn produce a Higgs boson. While the decay products of the Higgs boson are located in the central-rapidity range of the detector, the scattered quarks give rise to two jets in the far forward and backward regimes, thereby providing a very distinct event topology. Typically, these tagging jets are widely separated in rapidity and, due to the electroweak nature of the gauge boson exchange, additional QCD radiation is emitted preferentially close to one of the original partons in the hard process. This is different from the characteristics of the major QCD backgrounds, whose colour structur leads to widespread hadronic activity~\cite{Asner}.

\begin{figure}[t!]
\begin{center}
\begin{minipage}{32pc}
\includegraphics[width=32pc]{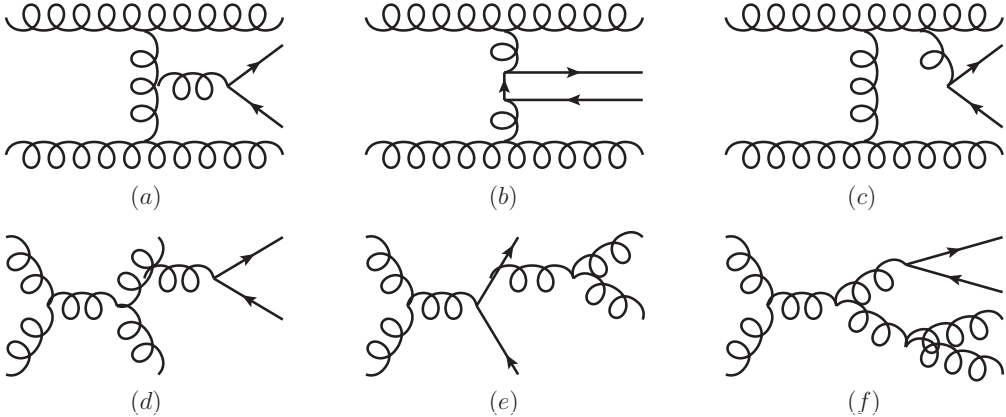}
\end{minipage}
\caption{\label{BGtopo}Typical Feynman diagrams for the dominant QCD-induced $b\bar{b}jj$ background. Diagrams with external quark lines instead of gluons are not shown.}
\end{center}
\end{figure}

If the Higgs boson exists, WBF will not only serve as a potential
discovery channel, but also allow to measure its mass, CP properties and
couplings to the top quark, tau lepton and weak gauge
bosons~\cite{Plehn:2001nj,Hankele:2006ma,Zeppenfeld:2000td,Duhrssen:2004cv}. However,
the determination of the $Hb\bar{b}$ Yukawa coupling remains
difficult~\cite{Mangano:2002wn}. The $H\rightarrow b\bar{b}$ decay
channel is overwhelmed by large QCD backgrounds which are dominated by
gluonic components (Fig.~\ref{BGtopo}). New strategies were proposed to
cope with these difficulties, such as exploring the jet substructure in
so-called ``fat jets''~\cite{fatJets}. Alternatively, requiring the
presence of a $W$ boson in association with the Higgs boson in WBF
processes can help to constrain the bottom quark Yukawa
coupling~\cite{Rainwater:2000fm,Ballestrero:2008kv}. The major drawback
of the extra heavy gauge boson emission is a significant decrease of the
cross section. This loss in statistics can be avoided to some extent by requiring the radiation of a photon instead of a $W$ boson~\cite{Gabrielli}. 

\begin{figure}[t!]
\begin{center}
\begin{minipage}{15pc}
\includegraphics[width=15pc]{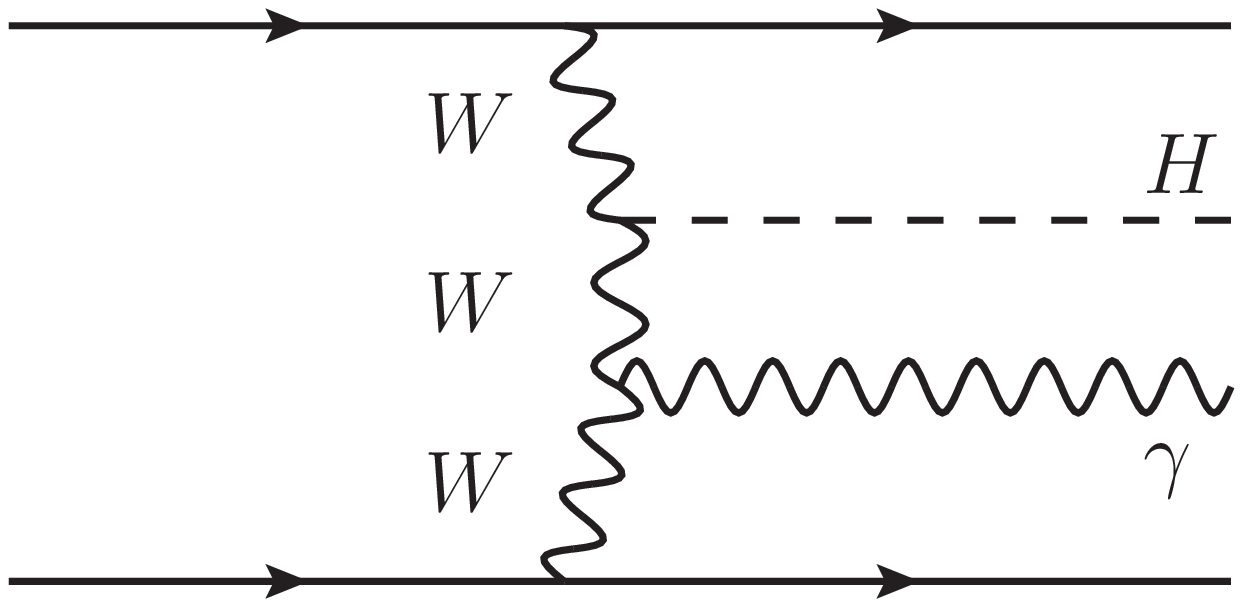}
\center\footnotesize $(a)$
\end{minipage}\hspace{4pc}%
\begin{minipage}{15pc}
\includegraphics[width=15pc]{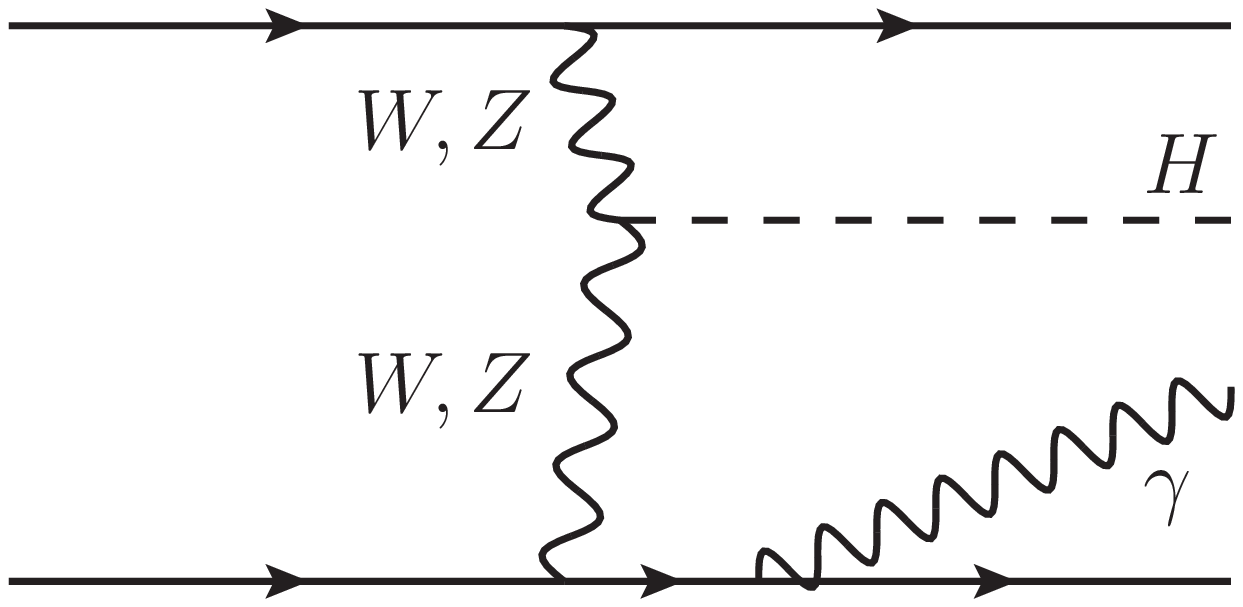}
\center\footnotesize $(b)$
\end{minipage}
\caption{\label{bornDiagrams}Typical Feynman diagrams contributing to $H\gamma jj$ production via WBF at LO.}
\end{center}
\end{figure}

Most importantly, the extra photon requirement gives rise to a
significant suppression of the gluon-dominated backgrounds, due to the absence of a gluon-to-photon coupling in the context of the Standard Model. 
Furthermore, in quark-scattering contributions that are connected by a
neutral $t$-channel gauge boson, destructive interference occurs between
diagrams emitting a photon from an incoming or an outgoing fermion line,
respectively. This effect arises in gluon-mediated QCD backgrounds as
well as in neutral current contributions to WBF processes, such as the
$Z$-exchange graphs of Fig.~\ref{bornDiagrams}~(b). Therefore, the
requirement of an additional hard central photon does not only suppress
QCD backgrounds, but also the $ZZ$ fusion component of the signal
channel, while the charged-current WBF contributions are relatively enhanced. 

A detailed analysis of the $H\gamma jj$ signal and its major backgrounds has been performed based on leading-order (LO) simulations in Ref.~\cite{Gabrielli}. In that work it has been shown that the central photon requirement reduces the cross section of the Higgs signal in WBF approximately by a factor of $1/100$, as expected from a naive estimate based on the size of the electromagnetic coupling constant. The relevant backgrounds instead drop by a factor of about $1/3000$, resulting in a statistical significance of $1\lessapprox S/\sqrt{B}\lessapprox 3$ for a luminosity of $100/$fb at a center-of-mass energy of $\sqrt{s}=14$~TeV and a Higgs mass of $m_H=120$~GeV. 
In a more recent parton shower study~\cite{Asner} a significance of the same order of magnitude has been obtained.

However, in order to obtain precise predictions accompanied by reliable estimates of their theoretical uncertainties, a next-to-leading order (NLO) QCD calculation of the signal process is essential. In~\cite{HAjj_AFJ}, we have provided  such a calculation within the parton level Monte-Carlo program \VBFNLO{}~\cite{VBFNLO,VBFNLOnew}, which is designed for the simulation of processes with electroweak bosons at hadron colliders.

\section{Details of the calculation}
\vspace{\baselineskip}
\subsection{Tree-level calculation and approximations}

WBF-induced Higgs boson production in association with a photon at hadron colliders mainly proceeds via quark-quark scattering processes, $qq\to qqH\gamma$, mediated by the exchange of a massive weak boson in the $t$-channel. The Higgs boson is radiated off this weak boson, while the photon can be emitted either from a fermion line or from a $W$~boson in the $t$-channel. Representative Feynman diagrams for the production process are depicted in Fig.~\ref{bornDiagrams}. The matrix elements for $H\gamma jj$ production are combined with a generic decay of the Higgs boson into a fermion-antifermion pair. NLO-QCD corrections are supplied for the production process only.

In order to speed up the calculation, we have used several approximations that have already successfully been applied to other WBF-type reactions. In our implementation, we disregard all diagrams of the type $q\bar{q}\rightarrow VH\gamma$ with a subsequent $V\rightarrow q\bar{q}$ decay, where $V=Z,W$. We also neglect interference terms between $t$- and $u$-channel diagrams. In the phase-space regions relevant for the Higgs search in the WBF channel, the disregarded contributions have been shown to be small~\cite{DennerDittmaier}.

\subsection{NLO-QCD corrections}

The virtual corrections comprise the interference of one-loop diagrams with the Born
amplitude. Due to color conservation, only selfenergy, vertex, and box corrections to either
the upper or the lower quark line need to be considered. Within our approximations, contributions from diagrams with a gluon connecting the upper and the lower quark line do not interfere with the corresponding Born diagrams. 

The combination of all virtual corrections to quark line $i$ gives rise to an amplitude of the form 
\begin{equation}
\mathcal{M}_V^{i}=\mathcal{M}_B \frac{\alpha_s(\mu_R)}{2\pi}C_F\left(\frac{4\pi\mu_R^2}{Q_i^2}\right)^\epsilon\Gamma(1+\epsilon)\left( -\frac{2}{\epsilon^2} - \frac{3}{\epsilon} + c_{\mathrm{virt}} \right) + \tilde{\mathcal{M}}_V^{i}\,,
\end{equation}
with $C_F=4/3$ and $c_{\mathrm{virt}}=\pi^2/3-7$ in dimensional reduction ($c_{\mathrm{virt}}=\pi^2/3-8$ in conventional dimensional regularization). For each quark line $i$, the quantity $Q_i$ is related to the momentum transfer of the incoming to the outgoing fermion with momenta $p_\mathrm{in}$ and $p_\mathrm{fi}$, respectively, by $Q_i^2=-(p_\mathrm{in}-p_\mathrm{fi})^2$.

\begin{figure}[t!]
\begin{center}
\begin{minipage}{15pc}
\includegraphics[width=15pc]{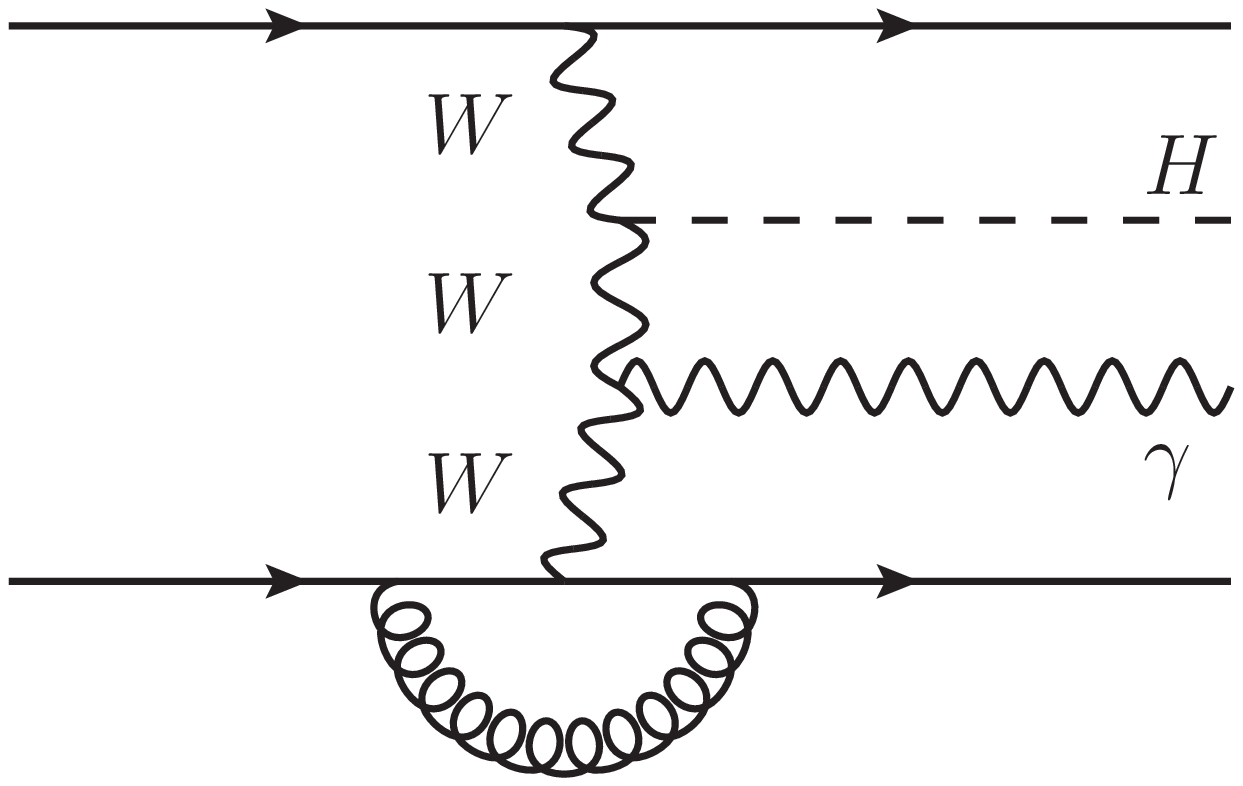}
\center\footnotesize $(a)$
\end{minipage}\hspace{4pc}%
\begin{minipage}{15pc}
\includegraphics[width=15pc]{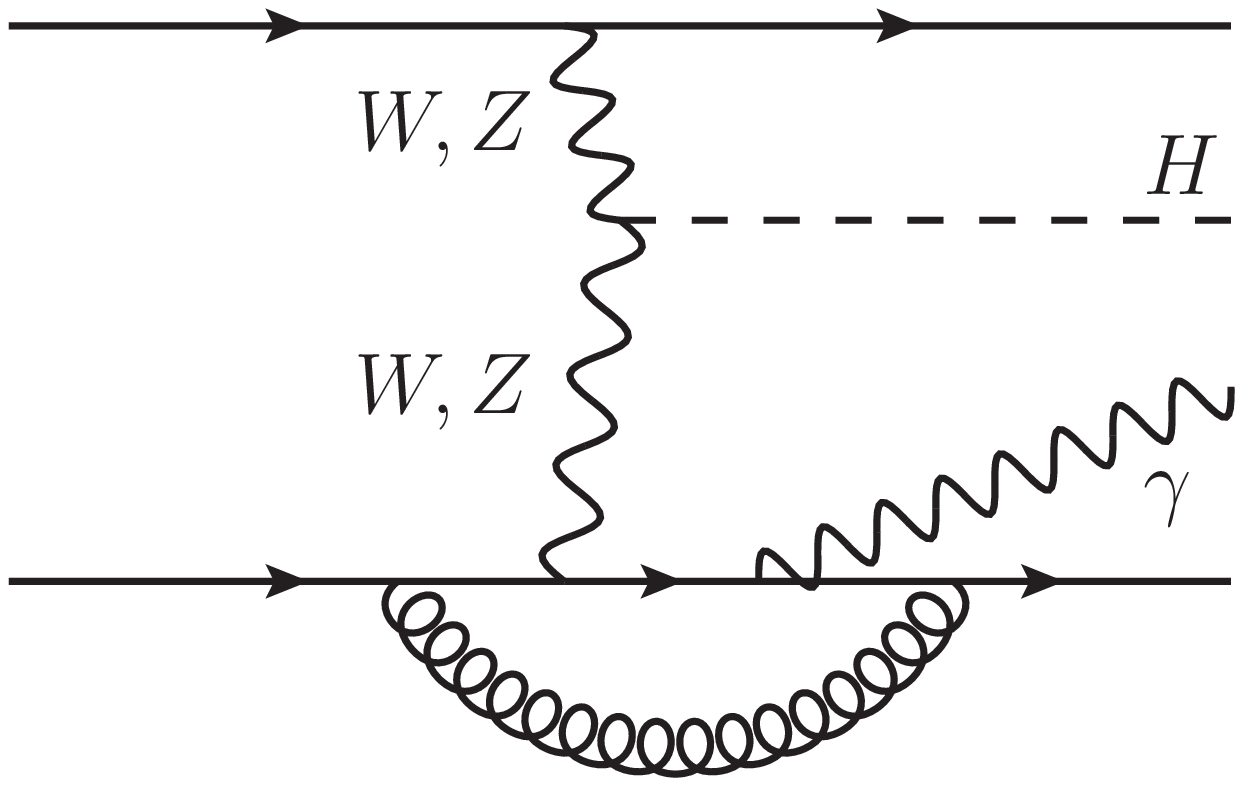}
\center\footnotesize $(b)$
\end{minipage}
\caption{\label{virtualDiagrams}Representative set of loop diagrams for $H\gamma jj$ production.}
\end{center}
\end{figure}

The finite parts $\tilde{\mathcal{M}}_V^{i}$ are evaluated numerically by means of a Passarino-Veltman tensor reduction~\cite{Passarino-Veltman}. 
For a small fraction of phase-space points, we encounter numerical
instabilities which can be traced back to small values of the emerging
Gram determinants. Such instabilities are monitored by testing Ward
identities at every phase-space point. The $\tilde{\mathcal{M}}_V^{i}$ contributions from phase-space points where these Ward identities are violated are removed. We have checked that the number of phase-space points that do not fulfill the Ward identities with a relative accuracy of at least $10^{-3}$ is less than 0.1 permille and that the numerical error due to disregarding these contributions is small. A detailed description of the calculation of the loop diagrams can be found in~\cite{Campanario}.

The real emission contributions are obtained from the LO diagrams by supplying all possible insertions of a gluon on the quark lines and considering crossing-related diagrams with a gluon in the initial state. We combine them with the virtual contributions by the dipole subtraction formalism of Catani and Seymour~\cite{Catani-Seymour}. 

\subsection{Checks}

Extensive checks have been made to verify our results. We have prepared
three independent implementations of the entire calculation, which agree
to better than one permille at the level of integrated cross sections.  
Furthermore, we have compared the leading order and real emission amplitudes with amplitudes generated by \texttt{MadGraph}~\cite{MadGraph} and found agreement to about 12 digits at a random set of phase-space points.
The leading order cross section was shown to be in accord with the results of \texttt{MadEvent}~\cite{MadEvent} within the given accuracy of the program. A precise comparison to the results of Ref.~\cite{Gabrielli} was difficult, as not all parameters of their calculation were explicitly listed in that publication. Nonetheless, we could reproduce their cross sections within a margin of $3-5\%$.

In addition, we have checked that the real emission amplitudes are invariant under QCD gauge transformations. All contributions obey QED gauge invariance. We furthermore validated our implementation of the dipole subtraction formalism by testing that the dipole terms are approaching the real emission contributions both in the soft and collinear limits.

\section{Results}
For the results that we are presenting here, we are assuming a proton-proton collider with a center-of-mass energy of $\sqrt{s}=14$ TeV. We are using the CTEQ6L1 parton distribution functions at LO and the CTEQ6M set with $\alpha_S(m_Z)=0.118$ at NLO~\cite{CTEQ}. As electroweak input parameters, 
\begin{align}
m_Z &= 91.188 \,\mathrm{GeV}\,, \nonumber\\
m_W &= 80.398 \,\mathrm{GeV}\,, \\
G_F &= 1.166 \times 10^{-5} \,\mathrm{GeV}^{-2}\,,\nonumber
\end{align}
are taken. 
Throughout, we have set the Higgs mass to $m_H=120$ GeV. Fermion masses are neglected and contributions with external top quarks are entirely disregarded. 

We are using the $k_T$ algorithm as described in~\cite{run2} with a cone
parameter of $R_C=0.7$ to define a jet. To avoid collinear singularities
associated with the final-state photon, we utilize the isolation
criterion introduced by Frixione~\cite{Frixione}. An event is considered acceptable if the hadronic energy deposited in a cone around the
photon is limited by
\begin{equation}
  \sum_{i:\Delta R_{i\gamma}<\Delta R} p_{Ti} \leq \frac{1-\cos\Delta
    R}{1 - \cos \delta_0} p_{T\gamma} \qquad\forall \Delta R
  \leq \delta_0.
\end{equation}
Here, the summation index $i$ runs over all final-state partons found in
a cone of size $\Delta R$ in the rapidity-azimuthal angle plane around the
photon, $p_{Ti}$ denotes the transverse momentum, and $\Delta
R_{i\gamma}$ the separation of parton $i$ from the photon, while $\delta_0$ stands for a fixed separation.
 
While we account for the kinematic distributions of the Higgs decay into two generic massless fermions (denoted by $b$ below, although not necessarily indicating a bottom quark), the branching ratio for this decay is not included in the results we are presenting here.

Our inclusive set of cuts is given by
\begin{align}
  \label{inclCuts}
  &\qquad\qquad&p_{Ti} &> 20 \,\text{GeV}\,, &  m_{jj}^{\text{tag}} &> 100 \,\text{GeV}\,,&\qquad\qquad&  \nonumber\\
  &\qquad\qquad&\lvert y_j \rvert &\le 5\,,& \lvert y_{\gamma,b}\rvert &\le 2.5\,,&\qquad\qquad& \\
  &\qquad\qquad&\Delta R_{ik} &\ge 0.4\,,&  \delta_0 &= 0.7\,,&\qquad\qquad& \nonumber
\end{align}
where $p_{Ti}$ stands for the transverse momentum of any jet, Higgs decay particle, or photon, and $\Delta R_{ik}$ is the separation in the rapidity-azimuthal angle plane of any two objects in the recombined final state.

\begin{table}[t!]
\caption{\label{Xsecs}Cross sections (in [fb]) within the WBF cuts of Eqs.\eqref{inclCuts}--\eqref{wbfCuts}, for different scale factors $\xi=\xi_R=\xi_F$. The statistical errors of the results are below the permille level and therefore not given here explicitly.
}

\begin{center}
\begin{tabular}{ccccc}
\br
$\xi$ & LO ($\mu_0^2=Q_i^2$) & LO ($\mu_0^2 = m_H^2 + \sum p_{Tj}^2$) & NLO ($\mu_0^2=Q_i^2$) & NLO ($\mu_0^2 = m_H^2 + \sum p_{Tj}^2$) \\
\mr
      $0.5$ & $15.72$ & $ 14.56 $ & $14.60$ & $14.84$\\
      $1.0$ & $14.65$ & $ 13.61 $ & $14.79$ & $14.84$\\
      $2.0$ & $13.70$ & $ 12.76 $ & $14.83$ & $14.75$\\
\br
\end{tabular}
\end{center}
\end{table}

Another, more stringent, set of cuts which will be entitled as ``WBF cuts'' in the following additionally comprises the typical cuts for Higgs boson searches in WBF. For this scenario, two tagging jets of large invariant mass have to be widely separated in rapidity in different hemispheres of the detector. The decay products of the Higgs boson are required to be located within the rapidity gap between the two tagging jets:
\begin{align}
  \label{wbfCuts}
  \Delta R_{ik} &\ge 0.7\,,  \nonumber\\
  m_{jj}^{\text{tag}} &> 600 \,\mathrm{GeV}\,,  \nonumber\\
  \lvert y_{j1} - y_{j2} \rvert &> 4\,, \\
  y_{j1} \cdot y_{j2} &< 0\,, \nonumber\\
  y_j^\mathrm{min} < &y_{\gamma,b} < y_j^\mathrm{max}\,. \nonumber
\end{align}

Numerical results for the integrated cross section at LO and NLO QCD within the latter set of cuts are shown in Tab.~\ref{Xsecs}. For the renormalization and factorization scales, 
\begin{align}
&\qquad\qquad&\mu_R = \xi_R\, \mu_0, && \mu_F = \xi_F\, \mu_0,&\qquad\qquad&
\end{align}
we employ two different central scales, defined by $\mu_0^2=Q_i^2$ and
$\mu_0^2 = m_H^2 + \sum p_{Tj}^2$, where the sum extends over defined jets. The scale factors, $\xi_R=\xi_F=\xi$, are varied by factors of two. Obviously, the dependence of the cross section on the factorization scale goes down when NLO-QCD corrections are included. The renormalization scale enters for the first time at NLO, since no strong coupling occurs in the tree-level process. 
For every scale choice, the ratio of the LO to the NLO cross section, referred to as K-factor, deviates from one by only a few percent.

\begin{figure}[t!]
\begin{center}
\begin{minipage}{30pc}
\includegraphics[width=30pc]{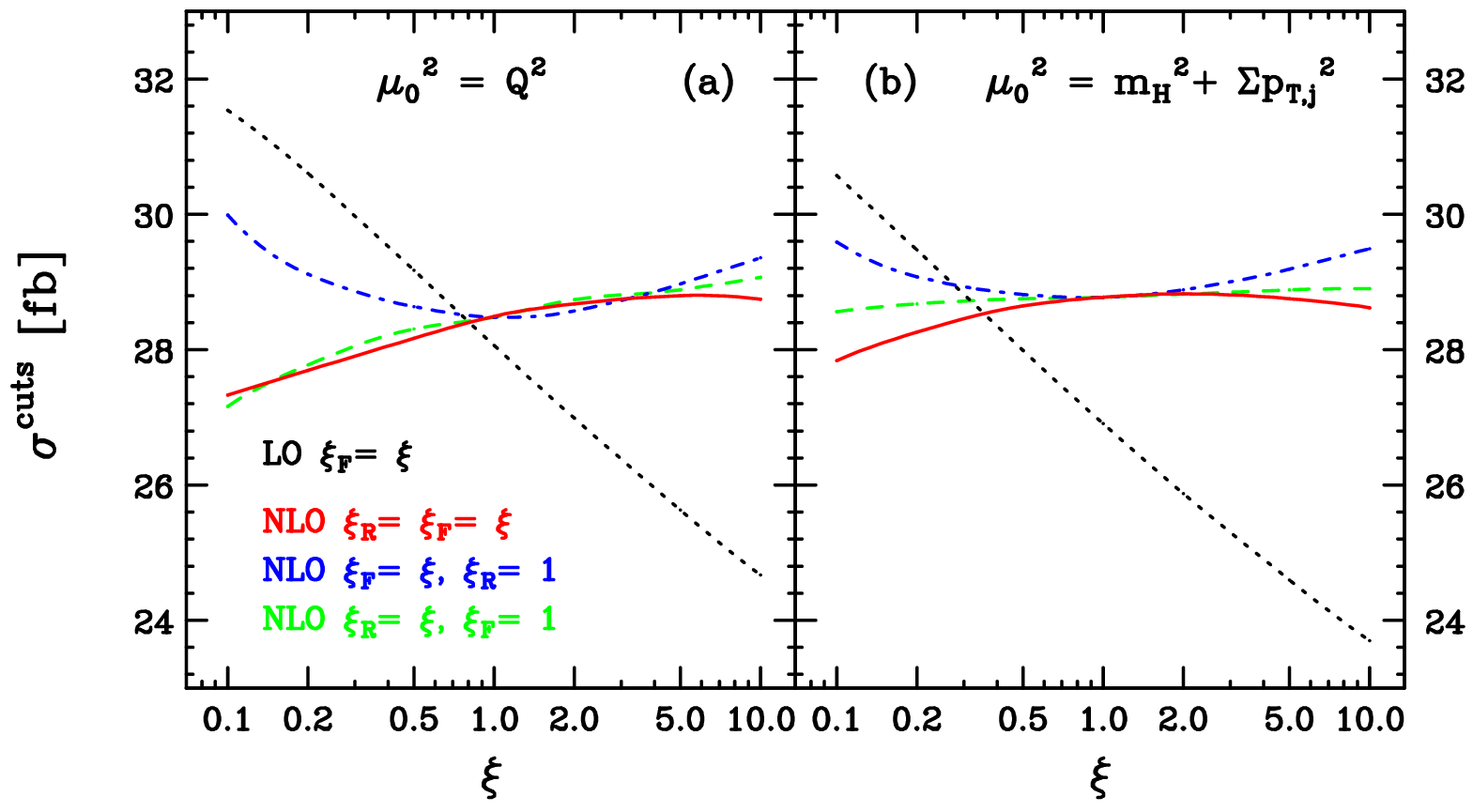}
\end{minipage}
\caption{\label{scaleDeps}
Scale dependence of the integrated cross section within the inclusive cuts of 
Eq.~(\ref{inclCuts}) at LO and NLO for two different choices of $\mu_{0}$. Shown are curves for $\xi_F = \xi$ at LO (black dots), $\xi_R=\xi_F=\xi$ at NLO (red solid), $\xi_F=\xi$, $\xi_R=1$ at NLO (blue dash-dot), and $\xi_R=\xi$, $\xi_F=1$ at NLO (green dashes).}
\end{center}
\end{figure}

Figure~\ref{scaleDeps} illustrates the dependence of the integrated
cross section within our inclusive set of cuts, Eq.~\eqref{inclCuts}, on
the unphysical renormalization and factorization scales $\mu_R$ and
$\mu_F$. For $\mu_0^2=Q_i^2$ and $\mu_0^2=m_H^2 + \sum p_{Tj}^2$ the
curves follow the same trend. In each case, the scale dependence
decreases drastically when NLO corrections are taken into account. In
the range of $0.5 \leq \xi_F = \xi_R \leq 2$, the value of the NLO cross
section varies by just $2\%$ for $\mu_0^2=Q_i^2$, and by $1\%$ for
$\mu_0^2=m_H^2 + \sum p_{Tj}^2$. Comparing the results of
Fig.~\ref{scaleDeps} with those of Tab.~\ref{Xsecs}, we find that the
WBF cuts of Eq.~\eqref{wbfCuts} reduce the signal by about a factor of two.

\begin{figure}[t!]
\begin{center}
\begin{minipage}{30pc}
\includegraphics[width=30pc]{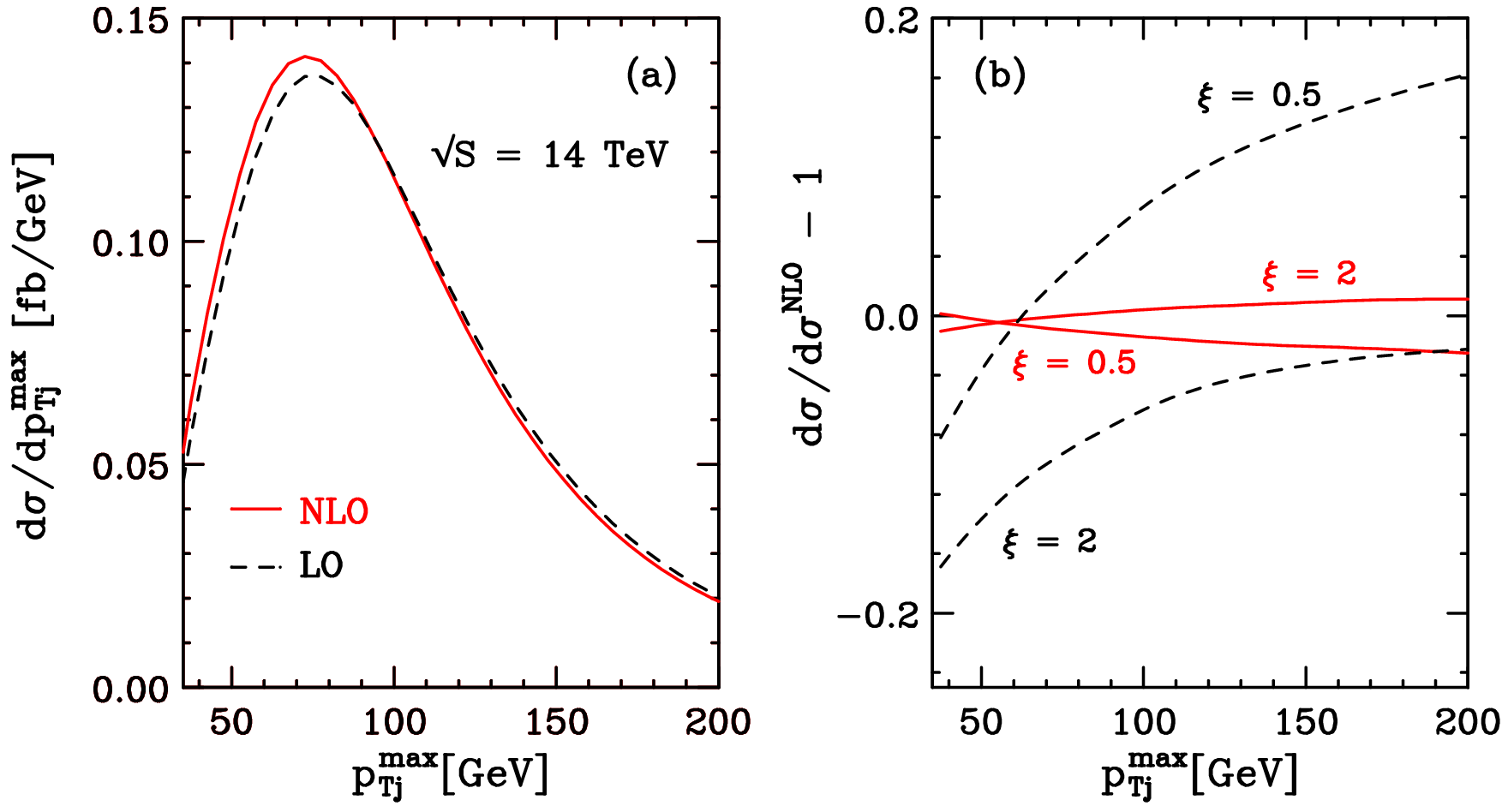}
\end{minipage}
\caption{\label{ptTag}
Transverse momentum distribution of the hardest tagging jet at LO (dashed black line) and NLO (solid red line) [panel~(a)] and relative corrections according to Eq.~\eqref{relvar}, when the factorization and renormalization scales are varied in the range $Q_i/2\leq\mu_R=\mu_F\leq 2Q_i$ [panel~(b)]. The WBF cuts according to Eqs.\eqref{inclCuts} and \eqref{wbfCuts} are employed here.}
\end{center}
\end{figure}

For the selection of WBF events in a collider environment, understanding the kinematic properties of the tagging jets is of vital importance. Fig.~\ref{ptTag}~(a) shows the distribution of the hardest tagging jet at LO and NLO within the WBF cuts of Eqs.~\eqref{inclCuts}--\eqref{wbfCuts}. While the NLO corrections are small in total, the shape of the transverse-momentum distribution changes significantly when going from LO to NLO,  with the peak of the distribution being shifted to slightly smaller values. 
In order to quantify the impact of the NLO-QCD corrections on the distribution of an observable $\mathcal{O}$ together with the scale uncertainties of the LO and the NLO prediction, we introduce the relative variation 
\begin{align}
\label{relvar}
\delta(\mathcal{O})=\frac{\mathrm{d}\sigma(\xi)/\mathrm{d}\mathcal{O}}{\mathrm{d}\sigma^{\mathrm{NLO}}(\xi=1)/\mathrm{d}\mathcal{O}}-1\,,
\end{align}
where $d\sigma(\xi)/d\mathcal{O}$ denotes the LO or NLO expression, evaluated for arbitrary values of the scale parameter $\xi=\xi_F=\xi_R$. The choice of $\mu_0$ is identical for $d\sigma/d\mathcal{O}$ and $d\sigma^\mathrm{NLO}/d\mathcal{O}$. 
For the transverse momentum distribution, this relative correction is displayed in Fig.~\ref{ptTag}~(b) for two different values of $\xi$. The bands in between the curves for $\xi=2$ and $\xi=0.5$ indicate the scale uncertainties of the LO and the NLO predictions, respectively. At NLO, this uncertainty is smaller than $4\%$ over the entire $p_T$ range considered. The scale dependence is much more pronounced at LO, though, and is generally larger for higher transverse momenta. 

In order to distinguish the Higgs signal in the $H\to b\bar b$ decay mode from QCD-induced $b\bar b (\gamma)$ backgrounds, imposing a cut on the invariant mass of the $b\bar{b}\gamma$ system can help~\cite{Gabrielli}. 
\begin{figure}[t!]
\begin{center}
\begin{minipage}{30pc}
\includegraphics[width=30pc]{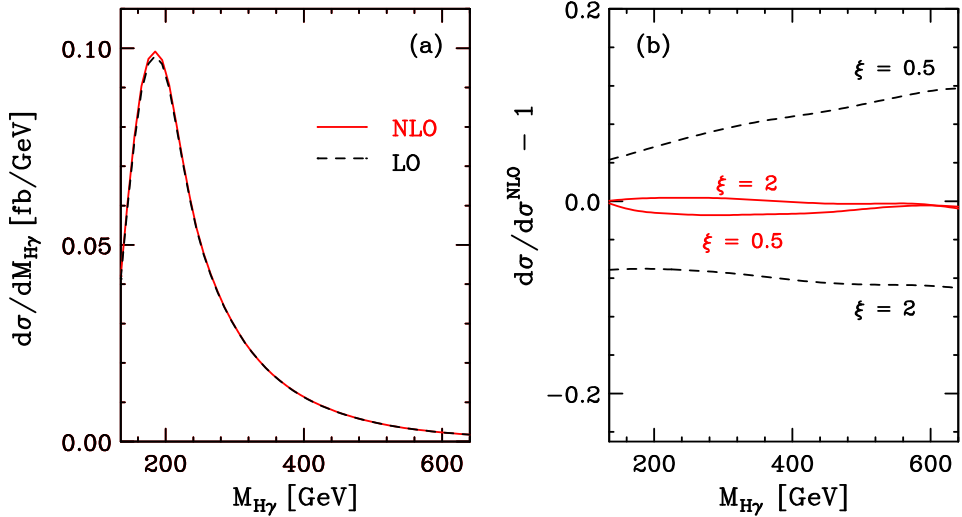}
\end{minipage}
\caption{\label{mhaGab}
Invariant mass distribution of the Higgs boson-plus-photon system at LO (dashed black line) and NLO (solid red line) [panel~(a)] and relative corrections according to Eq.~\eqref{relvar}, when the factorization and renormalization scales are varied in the range $Q_i/2\leq\mu_R=\mu_F\leq 2Q_i$ [panel~(b)].}
\end{center}
\end{figure}
Figure~\ref{mhaGab} illustrates the perturbative uncertainties associated with the invariant mass distribution of the Higgs boson-plus-photon system, being reconstructed from the four-momenta of the photon and the decay products of the Higgs boson. The distribution vanishes for $M_{H\gamma}<m_H$ and peaks for $m_H=120$~GeV at around $M_{H\gamma}\sim 165$~GeV. For larger Higgs masses, $d\sigma/dM_{H\gamma}$ would be shifted to correspondingly higher values. 

\section{Summary}
We have presented an NLO-QCD calculation for Higgs production in association with a photon via weak boson fusion at the LHC in the form of a fully flexible parton-level Monte Carlo program. Our calculation has been implemented in the publicly available \VBFNLO{}~package~\cite{VBFNLO,VBFNLOnew}.

Analyzing the $pp\to H\gamma jj$ process for two different combinations
of selection cuts, we have found that the impact of the NLO-QCD
corrections on the integrated cross section is generally small. Its
actual numerical value slightly depends on the choice of renormalization
and factorization scales. The shape of some distributions changes,
however, beyond leading order. For observables and regions of
phase-space within typical WBF cuts, the NLO corrections are mostly
smaller than $15\%$.

\ack{
We would like to thank F.~Campanario for useful discussions. 
This work was supported in part by the Deutsche Forschungsgemeinschaft
under SFB TR-9 ``Computational Particle Physics'' and via the 
Graduiertenkolleg ``High Energy Physics and Particle Astrophysics'', and by the Initiative and Networking Fund of the Helmholtz Association, contract HA-101 ("Physics at the Terascale"). 
T.~Figy would like to thank the Institute for Particle Physics Phenomenology at Durham University and the CERN Theory Division for their support. }

\section*{References}
\bibliography{ref}

\end{document}